\documentclass[12pt]{article}
\usepackage{amsmath, amsthm}
\usepackage{amscd}
\usepackage{amssymb}
\usepackage{array}
\usepackage{color}
\usepackage[margins]{trackchanges}

\newtheorem{thm}{Theorem}[section]
\newtheorem{theorem}[thm]{Theorem}

\newtheorem{lemma}[thm]{Lemma}
\newtheorem{proposition}[thm]{Proposition}

\theoremstyle{definition}
\newtheorem{definition}[thm]{Definition}

\begin{document}

\newcommand{\comment}[1]{{\color{blue}\rule[-0.5ex]{2pt}{2.5ex}}
\marginpar{\scriptsize\begin{flushleft}\color{blue}#1\end{flushleft}}}

\newcommand{\be}{\begin{equation}}
\newcommand{\ee}{\end{equation}}
\newcommand{\beq}{\begin{equation}}
\newcommand{\eeq}{\end{equation}}
\newcommand{\ba}{\begin{align}}
\newcommand{\ea}{\end{align}}
\newcommand{\ban}{\begin{align*}}
\newcommand{\ean}{\end{align*}}

\newcommand{\id}{\relax{\rm 1\kern-.28em 1}}
\newcommand{\R}{\mathbb{R}}
\newcommand{\N}{\mathbb{N}}
\newcommand{\C}{\mathbb{C}}
\newcommand{\Z}{\mathbb{Z}}
\newcommand{\g}{\mathfrak{G}}
\newcommand{\e}{\epsilon}

\newcommand{\hs}{\hfill\square}
\newcommand{\hbs}{\hfill\blacksquare}

\newcommand{\bp}{\mathbf{p}}
\newcommand{\bmax}{\mathbf{m}}
\newcommand{\bT}{\mathbf{T}}
\newcommand{\bU}{\mathbf{U}}
\newcommand{\bP}{\mathbf{P}}
\newcommand{\bA}{\mathbf{A}}
\newcommand{\bm}{\mathbf{m}}
\newcommand{\bIP}{\mathbf{I_P}}

\newcommand{\cA}{\mathcal{A}}
\newcommand{\cB}{\mathcal{B}}
\newcommand{\cC}{\mathcal{C}}
\newcommand{\cI}{\mathcal{I}}
\newcommand{\cO}{\mathcal{O}}
\newcommand{\cG}{\mathcal{G}}
\newcommand{\cJ}{\mathcal{J}}
\newcommand{\cF}{\mathcal{F}}
\newcommand{\cP}{\mathcal{P}}
\newcommand{\ep}{\mathcal{E}}
\newcommand{\E}{\mathcal{E}}
\newcommand{\cH}{\mathcal{O}}
\newcommand{\cPO}{\mathcal{PO}}
\newcommand{\cl}{\ell}
\newcommand{\cFG}{\mathcal{F}_{\mathrm{G}}}
\newcommand{\cHG}{\mathcal{H}_{\mathrm{G}}}
\newcommand{\Gal}{G_{\mathrm{al}}}
\newcommand{\cQ}{G_{\mathcal{Q}}}
\newcommand{\cT}{\mathcal{T}}
\newcommand{\cM}{\mathcal{M}}

\newcommand{\ri}{\mathrm{i}}
\newcommand{\re}{\mathrm{e}}
\newcommand{\rd}{\mathrm{d}}
\newcommand{\rSt}{\mathrm{St}}
\newcommand{\rGL}{\mathrm{GL}}
\newcommand{\rSU}{\mathrm{SU}}
\newcommand{\rSL}{\mathrm{SL}}
\newcommand{\rSO}{\mathrm{SO}}
\newcommand{\rOSp}{\mathrm{OSp}}
\newcommand{\rSpin}{\mathrm{Spin}}
\newcommand{\rsl}{\mathrm{sl}}
\newcommand{\rM}{\mathrm{M}}
\newcommand{\rU}{\mathrm{U}}
\newcommand{\rdiag}{\mathrm{diag}}
\newcommand{\rP}{\mathrm{P}}
\newcommand{\rdeg}{\mathrm{deg}}
\newcommand{\rStab}{\mathrm{Stab}}
\newcommand{\rcof}{\mathrm{cof}}

\newcommand{\M}{\mathrm{M}}
\newcommand{\End}{\mathrm{End}}
\newcommand{\Hom}{\mathrm{Hom}}
\newcommand{\diag}{\mathrm{diag}}
\newcommand{\rspan}{\mathrm{span}}
\newcommand{\rank}{\mathrm{rank}}
\newcommand{\Gr}{\mathrm{Gr}}
\newcommand{\ber}{\mathrm{Ber}}

\newcommand{\fsl}{\mathfrak{sl}}
\newcommand{\fg}{\mathfrak{g}}
\newcommand{\ff}{\mathfrak{f}}
\newcommand{\fgl}{\mathfrak{gl}}
\newcommand{\fosp}{\mathfrak{osp}}
\newcommand{\fm}{\mathfrak{m}}

\newcommand{\ttau}{\tilde\tau}

\newcommand{\str}{\mathrm{str}}
\newcommand{\Sym}{\mathrm{Sym}}
\newcommand{\tr}{\mathrm{tr}}
\newcommand{\defi}{\mathrm{def}}
\newcommand{\Ber}{\mathrm{Ber}}
\newcommand{\spec}{\mathrm{Spec}}
\newcommand{\sschemes}{\mathrm{(sschemes)}}
\newcommand{\sschemeaff}{\mathrm{ {( {sschemes}_{\mathrm{aff}} )} }}
\newcommand{\rings}{\mathrm{(rings)}}
\newcommand{\Top}{\mathrm{Top}}
\newcommand{\sarf}{ \mathrm{ {( {salg}_{rf} )} }}
\newcommand{\arf}{\mathrm{ {( {alg}_{rf} )} }}
\newcommand{\odd}{\mathrm{odd}}
\newcommand{\alg}{\mathrm{(alg)}}
\newcommand{\sa}{\mathrm{(salg)}}
\newcommand{\sets}{\mathrm{(sets)}}
\newcommand{\SA}{\mathrm{(salg)}}
\newcommand{\salg}{\mathrm{(salg)}}
\newcommand{\varaff}{ \mathrm{ {( {var}_{\mathrm{aff}} )} } }
\newcommand{\svaraff}{\mathrm{ {( {svar}_{\mathrm{aff}} )}  }}
\newcommand{\ad}{\mathrm{ad}}
\newcommand{\Ad}{\mathrm{Ad}}
\newcommand{\pol}{\mathrm{Pol}}
\newcommand{\Lie}{\mathrm{Lie}}
\newcommand{\Proj}{\mathrm{Proj}}
\newcommand{\rGr}{\mathrm{Gr}}
\newcommand{\rFl}{\mathrm{Fl}}
\newcommand{\rPol}{\mathrm{Pol}}
\newcommand{\rdef}{\mathrm{def}}
\newcommand{\rE}{\mathrm{E}}

\newcommand{\sym}{\cong}
\newcommand{\al}{\alpha}
\newcommand{\lam}{\lambda}
\newcommand{\de}{\delta}
\newcommand{\D}{\Delta}
\newcommand{\s}{\sigma}
\newcommand{\lra}{\longrightarrow}
\newcommand{\ga}{\gamma}
\newcommand{\ra}{\rightarrow}

\newcommand{\tit}{t}
\newcommand{\ts}{\tilde{s}}
\newcommand{\ty}{\tilde{y}}
\newcommand{\titp}{{t}^{\, '}}
\newcommand{\hv}{ \hat{t}}
\newcommand{\hu}{\hat{\tau}}
\newcommand{\hy}{\hat{y}}
\newcommand{\hS}{\hat{S}(x)}
\newcommand{\hx}{\hat{x}}
\newcommand{\hT}{\hat{T}}

\newcommand{\NOTE}{\bigskip\hrule\medskip}

\smallskip
    \centerline{\LARGE \bf  Quantum Supertwistors }

\vskip 2cm
\centerline{R. Fioresi$^1$,  M. A. Lled\'{o}$^{2}$} \vskip 1cm

\centerline{\it $^1$ Dipartimento di Matematica, Piazza Porta San Donato, 5 and}

\centerline{\it FaBiT, Universit\`{a} di
Bologna, 40126 Bologna, Italy}
\bigskip

\centerline{\it $^2$
Departament de F\'{\i}sica Te\`{o}rica, Universitat de Val\`{e}ncia and}

\centerline{\it IFIC(CSIC-UVEG)}\centerline{\it  C/ Dr. Moliner, 50, 46100 Burjassot.  Spain.}

\bigskip

\centerline{{\footnotesize e-mail: fioresi@dm.unibo.it,
Maria.Lledo@ific.uv.es}}

\vskip 1cm

\begin{abstract}
In this paper we give an explicit expression  for a star product on the super Minkowski space written in the supertwistor formalism. The big cell of the super Grassmannian  $\Gr(2|0, 4|1)$ is identified with the chiral, super Minkowki space. The super Grassmannian is an homogeneous space under the action of the complexification  $\rSL(4|1)$ of $\rSU(2,2|1)$, the superconformal group in dimension 4, signature (1,3) and supersymmetry $N=1$. The quantization is done by substituting the groups and homogeneous spaces by their quantum deformed counterparts. The calculations are done in Manin's formalism. When we restrict to the big cell we can compute explicitly an expression for the super star product in the  Minkowski superspace associated to this deformation and the choice of a certain basis of monomials.
\end{abstract}

\vskip1cm
Key words: star products, superspace, non commutative spacetime,
quantum groups, quantum supergroups.

\newpage
\section{Introduction}
Twistor theory  \cite{pe,pm,ww} {was initiated by Penrose as an alternative way of describing spacetime.} One starts with an abstract four dimensional complex vector space (twistor space) and the complex, compactified Minkowski space is  the set of two planes inside the twistor space. This is the Grassmannian manifold $G(2,4)$,  which is a homogeneous space of the group $\rSL(4,\C)$,
that is, an homogeneous space of the group $\rSL(4,\C)$. This group is the spin group of the conformal group of spacetime, namely $\rSU(2,2)$.

One passes from the Minkowski space to conformal space by a  compactification and vice versa by restricting to the  big cell of the conformal space. So one could think of a non conformally symmetric field theory as a conformal theory broken down to the big cell by some extra terms.

Conformal symmetry has a  fundamental role in the gauge/ gravity correspondence \cite{mal1} (for a  review see   \cite{mal2,ae}), which relates gravity theories to {conformal} gauge theories defined on {the} boundary of spacetime.
{It would then be interesting to see how conformal theories can be deformed and what is the meaning of the deformation from the  gravity point of view.}

In the original papers \cite{pe,pm}, Penrose believed that twistor theory could help to introduce the indetermination principle in spacetime. The  points had to be `smeared out' since in this formalism a point of spacetime is not a fundamental quantity, but it is secondary to twistors.

Nevertheless, all the twistor construction is classical. Our point of view is
 deforming the algebra of functions over spacetime to a noncommutative algebra.
Because of the non commutativity, this will introduce an indetermination principle among the coordinates of spacetime.
A quantum group
is a commutative but  non cocommutative  Hopf algebra, depending on an indeterminate parameter $q$. One can specify $q=1$ to recover the original commutative Hopf algebra, which is just a Lie group, or to any real or complex value to obtain examples of non commutative Hopf algebras.

The quantum group   $\rSL_q(4,\C)$ {is the} quantum conformal group once complexified. The idea underlying the work of \cite{fi2,fi3} was to make such  substitution and then to obtain a quantum Grassmannian, a quantum Minkowski space and a quantum Poincar\'{e} group satisfying the same relations among them as their classical counterparts. So the quantum conformal group acts naturally on the quantum Grassmannian, viewed as a {quantum homogeneous space}, and  the quantum Poincar\'{e} group is identified with the subgroup of it that preserves the big cell.  This construction has also been generalized to flag manifolds \cite{fi4}.

In the super setting, we have several superspaces that are of interest: the Grassmannian supervariety $\rGr(2|0,4|1)$, which corresponds in physical terms to the superalgebra of  chiral superfields and the superflag $\rFl(2|0, 2|1,4|1)$ which is the complexification of the $N=1$  Minkowski superspace. The same idea can be applied here with the supergroup $\rSL(4|1)$ \cite{cfl1,cfl2}, which also can be deformed to a quantum supergroup. For a detailed treatment of all, the super and  non super, classical and quantum  cases see \cite{fl3}. We will follow Manin's formalism \cite{ma} for quantum supergroups.

Here we deal with both cases, the super and non super one. We have identified a quantization of the (super)conformal space  as an homogeneous space of $\rSL_q(4|1)$.  In the big cell (the (super)Minkowski) it  can be presented as a {concrete} star product on the algebra of functions. There is an atlas of the Grassmannian with 6 identical cells, and the (super) star products in the intersections glue in such way that one can recover the quantum Grassmannian.

{Any conformal theory expressed in terms of twistors would have presumably, with this procedure,  a quantum counterpart, where the word `quantum' here means that we are deforming spacetime itself. The observables of such theories will be modified, but to compute the modifications explicitly it is not enough to look at the abstract algebra defined by generators and relations. One has to go to a `semiclassical' approach where observables are still  functions on spacetime as the original ones,  but with a product that is non commutative. This is achieved only if one has an explicit expression for the star product. In fact, the functions become (formal) series in the parameter of the deformation and there is no canonical choice for the star product among the equivalent ones. This problem is  present in general in deformation quantization. For example, when one quantizes  the phase space with a constant Poisson bracket, there seems to be a choice in which the star product has an easy computable expression,  the Moyal-Weyl star product. Other than this simple case there are not, to the best of our knowledge, computations showing explicitly different star products, less being quadratic (at the leading order) deformations. To compute the star product one has to stick to a basis  in the abstract algebra, and the explicit expression for the star product depends on such choice. We have chosen here a sort of normal ordering, so the final  expression does not have symmetry between the entries of the star product as the Moyal-Weyl has. It would have been more difficult to show the existence of a basis where the star product appears to be more symmetric, and then the calculation of the star product would have been more involved. For this reason, we have sticken to a normal ordering.  There is one instance, though, were the ordering is not relevant and it is the deformation to order one in h ($q=\re^h$) of the commutator induced by the star product $f\star g-g\star f$. This is the Poisson bracket, that we also compute explicitly, with an expression far simpler than the star product itself. There is always an equivalent deformation where the Poisson bracket is the first term in the star product itself so there is no need to antisymmetrize. Knowing the Poisson bracket would then be useful to compute first order corrections to the theories at hand, induced by the non commutativity of spacetime.}

We first deal with the non super case \cite{cfln}.{We work} in the algebraic category, so we first give an explicit formula for the star product   among two polynomials in the big cell of the Grassmannian. Since the quantum algebras that we present here are deformations of the algebra of polynomials on Minkowski space, the star product that we obtain is {also} algebraic.

{In} the same reference \cite{cfln} it is  shown  that this deformation can be extended  to the set of smooth functions in terms of a differential star product. Since a differential operator is determined once it is given on polynomials, the bidifferential operators appearing in the star product are completely determined. The Poisson bracket {leading} the deformation is  a quadratic one, so the Poisson structure is neither symplectic nor regular. For the super case we obtain that, at least to the first order in $h$, the differentiability property is maintained.

 Examples of such  transition  from the category of algebraic varieties to the category of differential manifolds in the quantum theory are given in  \cite{fl1,ll,fll,fl2}.
  {There, the authors consider coadjoint orbits with the
Kirillov-Kostant-Souriau  symplectic form. }It was shown  that some algebraic star products do not have differential counterpart (not even modulo an equivalence transformation), so the results of \cite{cfln} are non trivial.  It is {remarkable} that one of the algebraic star products that does not have {a} differential extension   is the star product on the coadjoint orbits of $\rSU(2)$, associated to the standard quantization of angular momentum. For algebraic star products and their classification, see also  \cite{ko2}.


 The Weyl-Moyal star product is, in some sense, the simplest formal deformation that one can construct on $\R^n$. It requires a constant Poisson bracket:
   $$\{f(x), g(x)\}= B^{\mu\nu}\partial_\mu f (x)\partial_\nu g (x),\qquad f, g\in  C^\infty(\R^n)\,,$$
where $B_{\mu\nu}$ is any constant, antisymmetric matrix. The associative, non commutative star product is given by \cite{we,moy}
 $$f\star g (x)=\sum_{n=0}^\infty \frac{h^n}{n!}B^{\mu_1\nu_1}\cdots
B^{\mu_n\nu_n}\partial_{\mu_1}\dots\partial_{\mu_n}f(x)
\partial_{\nu_1}\dots\partial_{\nu_n}g(x)\,.$$

 {In} \cite{ha},
  {the Minkowski space is endowed, first,  with a constant Poisson bracket as above.}
  {Then, using the R-matrix approach the authors construct the action of the conformal group on the noncommutative space, which gives a deformation similar to the one used in }
   \cite{kko}.

  { The Moyal deformation of space time has been used in string theory (the original references are} \cite{cds,sw}).
{ In string theory, the presence of a $B_{\mu\nu}$ field with a non zero vacuum expectation value can be interpreted as a deformation of space time with the Moyal-Weyl deformation induced by $\langle B_{\mu\nu}\rangle$. This is a genuine non commutative structure of spacetime. However, one has to take into account that it breaks the Lorentz invariance.}

\medskip

For the super case there are also this type of `{Moyal-Weyl}' deformation. It is known that the quantization of a Grassmann algebra is a Clifford algebra of split signature $(n,n)$ (see for example \cite{fell,lle}).

There are very few deformations that can be given explicitly as a star product in closed form \cite{bffls}. A general formula is known for an arbitrary Poisson bracket (Kontsevich's formula, \cite{ko}) but it is extremely hard to work out the coefficients for the differential operators appearing in the deformation, even for simpler, linear Poisson brackets. { For many deformations we only know how to express them in terms of generators and relations. While this maybe enough from a mathematical point of view, it is often not enough for applications. The formula that we give is involved but it is explicit, and this is a real advantage.}

 {In order to quantize the Grassmannian one can also use} the fact that the Grassmannian $G(m, n)\simeq \rSL(n)/P$, with $P$ a parabolic subgroup, is, as a real manifold, a coadjoint orbit of the group $\rSL(n)$. In fact, any flag manifold is so, being the  full flag $\rFl(1,2,3,\dots,n)$ the regular (maximal dimension) orbit and all the others non regular. The approaches of \cite{cg,cgr,cgr2,fl1,ll}  would then be relevant here. The Kirillov-Kostant-Soriau Poisson bracket on the coadjoint algebra given essentially by the Lie bracket, is a linear Poisson bracket. It restricts to a symplectic Poisson bracket on the orbits. {The} star product is obtained from the enveloping algebra {but it is only explicit once one takes symplectic coordinates on an open set of the orbit, in which case it is, locally, a Moyal-Weyl star product}. It is then {a} star product {equivariant} under the action of the group. In {the works mentioned above}, the quantization is given in terms of generators and relations so it is {an} algebraic {deformation}, but then in \cite{fll,fl2} the relation with differential star products was studied.
 This mechanism could, in principle, be extended to the super case.

 {Other works also deal with the quantization of spacetime in terms of the twistor space. A very interesting article is}
 \cite{ch}, which
 {applies the methods of geometric quantization to the twistor space. }

Another approach to the quantization of coadjoint orbits has been undertaken also in Refs. \cite{al,eem,mu} using the so-called  Shapovalov pairing  of Verma modules.

Grassmannians have also been quantized as  fuzzy spaces.  {A fuzzy space is built by using} harmonic functions on the coset space {and truncating} the expansion  at some level. {The functions can be expressed} as matrices {in a certain basis} and a product on the truncated space is defined just using matrix multiplication. We find this approach in \cite{dm,dj}.

We believe that the approaches just mentioned must be linked in some way, since the quantizations are equivariant under the classical group ($\rSL(4,\C)$ in this case) and all of them are intimately related to representation theory. It is, however, not straightforward to compare them.

\medskip

Interesting as these works are, our deformation is a different one. The Poisson bracket that we obtain on the  Minkowski (super)space is a quadratic one (in particular, not symplectic) and the star product is then non equivalent to a Weyl-Moyal one. Also, the equivariance of the star product is achieved only by deforming the group to a quantum group, contrary to the above mentioned approaches. Nevertheless, we are able to give an explicit formula for it in terms of a recursive expression. The formulas for the non super case are involved but manageable. For the super case,  we have put the star product in terms of the non super one, otherwise the notation becomes very heavy. This is an example of how a standard ordering in the generators of the quantum  Minkowski (super)space induce a (super) star product that it is not at all trivial. Contrary to many other deformations, whose algebra is given in terms of generators and relations, here we have an explicit (although involved) formula  for the calculation of the star product of two monomials.

\medskip

 The organization of the paper is as follows:

 In Section \ref{sec:Grassmannian} we review the classical picture, also for the super case,  and settle the notation for the algebraic approach. In Section \ref{superquantum-sec} we describe the quantum super Minkowski space obtained in Refs. \cite{fi2,fi3,cfl1, cfl2}, together with the corresponding quantum super groups. In Section \ref{even-sec} we tackle the even case studied in \cite{cfln} and give the explicit formula for the star product between two polynomials on Minkowski space. We refer to that same paper to see how  the differentiability of the star product is proven. Finally, in Section \ref{superstarproduct-section}, based on the results obtained in the previous section, we obtain the star product for the super Minkowski space

  For completeness, in Appendix \ref{diamond} we have given a basis of the super Poincar\'{e} group in terms of its usual generators.

\section{The classical chiral conformal and Minkowski superspaces}
\label{sec:Grassmannian}

We describe here the $N=1$ chiral conformal superspace  as
the super Grassmannian $\rGr:=\Gr(2|0, 4|1)$. The superspace $\rGr$
is an homogeneous superspace under the action of
supergroup $\rSL(4|1)$, which is
the complexification of the conformal supergroup in Minkowskian signature, namely $\rSU(2,2|1)$.
The chiral Minkowski superspace $\cM$ is realized as the big cell inside $\rGr$
and the action of the Poincar\'{e} supergroup as the symmetries
of $\rGr$ stabilizing $\cM$.

For an explicit construction of this picture, see
\cite{va,flv, ccf, cfl1, cfl2,  fl3}.
For the ordinary, non super counterpart description see also
\cite{ ww,cfln,va}. We briefly describe it here.

Let $\Gr_0=\rGr(2,4)$ denote the Grassmannian of $2$-dimensional subspaces
of $\C^4$. The space $\cT_0=\C^4$ is called the {\it twistor space}.
The Grassmannian
$\Gr_0=\rGr(2,4)$ is a complex analytic
manifold, a  projective algebraic
variety and an homogeneous space under the action of the group $\rSL(4,\C)$.  In fact, a 2-dimensional subspace is given by two independent vectors:
\beq\pi=(a,b)=\begin{pmatrix}a_1&b_1\\a_2&b_2\\a_3&b_3\\a_4&b_4\\\end{pmatrix}\label{vectors}\eeq
There is a natural right action of  $\rSL(2, \C)$ corresponding to
basis change.
The left action of $\rSL(4,\C)$ is the obvious one and it is a transitive
action. Selecting one element, say $\pi_0=(e_1, e_2)$ with
$\{e_i\}_{i=1}^4$ the canonical basis in $\C^4$, we find that the isotropy
group of $\pi_0$ is the upper parabolic subgroup
$$ P_0^u=\left\{\begin{pmatrix}L&M \\
0&R\end{pmatrix}\in \rSL(4,\C)\quad \big|\quad \det L \cdot \det R=1 \right\}\,,\label{ups0}$$ where $R$, $L$ and $M$ are $ 2\times 2$-matrices. So
$$\rGr_0=\rSL(4,\C)/P_0^u\,.$$

\medskip

The big cell of $\rGr_0$ is the set of points such that
\beq \det\begin{pmatrix}a_1&b_1\\a_2&b_2\end{pmatrix}\neq 0\,.\label{det0}\eeq
By a right $\rSL(2,\C)$ transformation we can bring $(\ref{vectors})$ to the standard form

\beq
U_{12}:=\left\{
\begin{pmatrix}1&0\\0&1\\t_{31}&t_{32}\\t_{41}&t_{42}\\\end{pmatrix}=
\begin{pmatrix}\id\\t\end{pmatrix}\right\}
\label{tunconstrained}\eeq with $t$ unconstrained. The big cell is then $U_{12}\approx \C^4$ . The subset of $\rSL(4,\C)$ leaving the big cell invariant is the lower parabolic subgroup

$$ P_0^l=\left\{\begin{pmatrix}x&0 \\
Tx&y\end{pmatrix}\in \rSL(4,\C)\quad \big|\quad \det x \cdot \det y=1 \right\}\,,\label{lps}$$  where the unconstrained matrix
$Tx$ is written in this way to see better the action on the big cell. We have
$$t\longrightarrow ytx^{-1} +T\,.$$ So $P_0^l$ is the Poincar\'{e} group including the dilations and the big cell is the Minkowski space, with its more familiar form  $t=x^\mu\sigma_\mu$ in terms of the Pauli matrices.

\medskip

As algebraic groups, the coordinate algebras of $\rSL_4:=\rSL(4,\C)$ and its subgroup $P_0^l$ are
\begin{align}
&\C[\rSL_4]= \C[g_{ij}]/(\det g-1), && i,j=1,\dots 4\\
&\C(P_0^l)=\C[x_{ij}, y_{kl}, T_{kj}]/(\det x\det y-1),  && i,j=1,2, \hbox{ and }k,l=3,4\,.
\end{align}
These algebras carry a well known commutative Hopf algebra structure.

\medskip

We can associate with $\Gr_0$ the $\Z$-graded ordinary algebra
$\cO(\Gr_0)$ given by its {\it Pl\"ucker embedding} in the projective space $\bP^5$ (see for example \cite{va, cfl1} or \cite{fl3} Ch. 2) i{ in terms of six indeterminates $q_{ij}$ and the Pl\"{u}cker relation described below}:
$$
\cO(\Gr_0):=\C[q_{ij}]/\cI_{P0}, \qquad 1\leq i<j \leq 4
$$
where $\cI_{P0}$ 
is the ideal generated by the Pl\"{u}cker relation:
\beq
q_{12}q_{34}-q_{13}q_{24}+q_{14}q_{23}=0\,.\label{plucker}
\eeq
The interesting observation here, that will be key to obtain the quantization, is that $\cO(\Gr_0)$
can be retrieved as a subalgebra of  $\C[\rSL_4]$. If we write the generators of $\C[\rSL_4]$
in its usual matrix form
\beq
g=\begin{pmatrix}g_{11}&g_{12}&g_{13}&g_{14}\\
g_{21}&g_{22}&g_{23}&g_{24}\\
g_{31}&g_{32}&g_{33}&g_{34}\\
g_{41}&g_{42}&g_{43}&g_{44}\end{pmatrix}\label{matrixgroup}\,,
\eeq
the determinants $d_{ij}=g_{i1}g_{j2}-g_{i2}g_{j1}$
with $1\leq i<j\leq 4$, that is, all possible determinants
of the two first columns,
satisfy the Pl\"{u}cker relation (\ref{plucker}) and
this is the only
independent relation that they satisfy (see \cite{fl3}). Therefore
$$
\cO(\Gr_0) \cong \C[d_{ij}]\subset\C[\rSL_4], \qquad 1\leq i<j\leq 4\,.
$$

The condition (\ref{det0}) is related to the invertibility of $q_{12}$ in (\ref{plucker}).
Introducing $q_{12}^{-1}$ with $q_{12}q_{12}^{-1}-1=0$ and degree $-1$, the subalgebra of
$\cO(\rGr_0)[q_{12}^{-1}]$ of degree 0 is the polynomial subalgebra freely generated
by the elements
$$
\begin{pmatrix}t_{31}&t_{32}\\t_{41}&t_{42}
\end{pmatrix}=
\begin{pmatrix}-d_{23}&d_{13}\\-d_{24}&d_{14}
\end{pmatrix}d_{12}^{-1}\,
$$
where $d_{ij}$ is the determinant formed by the elements of (\ref{matrixgroup}) in the positions columns 1,2 and rows $i, j$ ($i<j$).
The determinant $d_{34}$ can be obtained from (\ref{plucker}).
The calculation follows by taking the first two columns in (\ref{matrixgroup}) and multiplying on the right by
$$\begin{pmatrix}g_{11}&g_{12}\\g_{21}&g_{22}\end{pmatrix}^{-1}=\frac 1{d_{12}}\begin{pmatrix}g_{22}&-g_{12}\\-g_{21}&g_{11}\end{pmatrix}\,,$$
and then comparing with (\ref{tunconstrained}).

\bigskip
Similarly we consider the $N=1$ {\it  supertwistor superspace},  $\C^{4|1}$. We have  the set
$\rGr:=\rGr(2|0,4|1)$ of $2|0$ subspaces in $\C^{4|1}$.
It is naturally
an analytic supermanifold, a projective algebraic supervariety and an homogeneous superspace under
the action of the supergroup $\rSL(4|1)$. We will use the language of the {\it functor of points}
(see for example \cite{ccf, demo, fl3}), better suited to treat supergroups. So one has
\beq
\rSL(4|1)(\cA)=\left\{\begin{pmatrix}g_{ij}&\gamma_{i5}\\\gamma_{5j}& g_{55}\end{pmatrix},
\quad i,j=1,\dots 4\right\}\label{sl41}
\eeq
where $\cA$ is any
superalgebra.
The Latin letters will represent elements of $\cA_0$ and the Greek  ones elements of $\cA_1$
unless otherwise stated.

 We can give an element in $\rGr(\cA)$,
for $\cA$ local, in terms of two even independent vectors
\beq\pi=(a,b)=\begin{pmatrix}a_1&b_1\\a_2&b_2\\a_3&b_3\\a_4&b_4\\\alpha_5&\beta_5\end{pmatrix}
\label{svectors}\eeq that, as before, can be chosen up to the right action of $\rSL_2(\cA)$.
The isotropy group of $\{e_0,e_1\}$, being $\{e_i, \epsilon_5,\,\, i=1,\dots 4\}$
the canonical basis, is
 the upper parabolic subgroup
$$ P^u(A)=\left\{\begin{pmatrix}L&M&\alpha \\
0&R&\beta\\
0&\delta&d\end{pmatrix}\in \rSL(4|1)(\cA)\quad \big|\quad \det L \cdot \det R=d \right\}\,,\label{ups}$$ so
$$\rGr(\cA)=\rSL(4|1)(\cA)/P^u(A)\,.$$
 The description of homogeneous spaces for super Lie groups is done in detail in \cite{flv}. We will say that $\rGr$ is the $N=1$ {\it chiral superspace}.

 The reduced manifold of $\rGr$ is $\rGr_0$, so the big cell of $\rGr (\cA)$ will be the set of $\cA$-points with
\beq
\det\begin{pmatrix}a_1&b_1\\a_2&b_2\end{pmatrix}
\, \hbox{invertible} \label{det}\eeq
which with a right $\rSL_2(\cA)$ 
transformation can be brought to the standard form
 \beq\begin{pmatrix}1&0\\0&1\\t_{31}&t_{32}\\t_{41}&t_{42}\\
 \tau_{51}&\tau_{52}\end{pmatrix}=\begin{pmatrix}\id\\t\\
 \tau\end{pmatrix}\,.\label{supertwistorspace}\eeq
We call the set of such matrices the big cell $U_{12}$; this is an open
subsupermanifold of $\Gr$ and $U_{12} \cong \C^{4|2}$.
The subgroup leaving invariant the big cell is the lower parabolic
subgroup of  $\rSL(4|1)$:
 \beq P^l=\begin{pmatrix}x&0&0\\
 Tx&y& y\eta\\ d\rho&d\xi&d\end{pmatrix}\label{lpss}\eeq and the action  on the big cell is
 $$\begin{pmatrix}\id\\t'\\
 \tau'\end{pmatrix}=\begin{pmatrix}\id\\y(t+\eta\tau)x^{-1}+T\\d(\tau+\rho+\xi t)x^{-1}
 \end{pmatrix}\,.$$
 Taking  $\xi=0$ we obtain the super Poincar\'{e} group including dilations. The condition $\xi=0$ will be necessary if we also consider the antichiral superspace $\rGr(2|1,4|1)$.

\medskip

There is a $\Z$-graded superalgebra
associated with the  super Pl\"ucker  embedding of $\rGr$ in $\bP^{6|4}$ (see \cite{cfl1}, \cite{fl3} Ch. 4 for more details) {given in terms of seven even indeterminates $q_{ij}$, $a_{55}$ and four odd indeterminates satisfying the relations described below}:
\beq
\cO({\Gr}):=\C[q_{ij}, \lam_k, a_{55}]/\cI_P, \qquad i,j, k=1,\dots, 4\,, \label{sgr}
\eeq
where $I_{P}$ is the super ideal generated by the
super Pl\"{u}cker relations:
\begin{align}
&q_{12}q_{34}-q_{13}q_{24}+q_{14}q_{23}=0, &&
\hbox{ (classical Pl\"{u}cker relation)} \nonumber  \\&q_{ij}\lambda_k-q_{ik}\lambda_j+q_{jk}\lambda_i=0,&& 1\leq i<j<k\leq 4\nonumber \\
& \lambda_i \lambda_j=a_{55}q_{ij}&& 1\leq i<j\leq 4 \nonumber\\
& \lambda_ia_{55}=0\,.\label{superplucker} &&
\end{align}

As in the non super case, we consider the  first two rows in
(\ref{sl41}) and construct the quantities
$$d_{ij}=g_{i1}g_{j2}-g_{i2}g_{j1},\qquad \sigma_i=g_{1i}\gamma_{52}- g_{2i}\gamma_{51},\qquad a=\gamma_{51}\gamma_{52} $$ which satisfy the super Pl\"{u}cker relations and no other independent relations. In this way one retrieves the algebra $\C[q_{ij}, \lambda_i,a_{55}]/\cI_P$ of the super Pl\"{u}cker embedding as a subalgebra of $\rSL(4|1)$.


The condition (\ref{det}) defining the super Minkowski space as the big cell in $\rGr$ is then equivalent to $q_{12}\neq 0$.
The superalgebra of polynomials $\cO(\cM)$ on $\cM$ is
then retrieved in  $\cO({\Gr})$
 as the elements of degree zero in $\cO(\Gr)[q_{12}^{-1}]$.

This subalgebra, $\cO(U_{12})=\cO(\cM)$ is
the polynomial subalgebra generated by the elements
(see (\ref{supertwistorspace}))

\beq\begin{pmatrix}
t_{31}&t_{32}\\t_{41}&t_{42}\nonumber\\
\tau_{51}&\tau_{52}
\end{pmatrix}=
\begin{pmatrix}
-d_{23}&d_{13}\\
-d_{24}&d_{14}\\
\sigma_1&\sigma_2
\end{pmatrix}d_{12}^{-1}
\label{minkowskiss}
\eeq
and the rest of the indeterminates can be obtained from these using the super Pl\"{u}cker relations (\ref{superplucker}).
(\cite{fl3} Ch. 5).

As supergroups, $\rSL(4|1)$, $P^l$ and also $P^u$ are super Hopf algebras. This  allows us to give the coaction of the relevant supergroup on the super conformal or super Minkowski spaces.

\section{The quantum super Minkowski space}\label{superquantum-sec} We pass now to describe briefly the quantization of super Minkowski space obtained in \cite{cfl1}, Sec. 2 (see also in \cite{fl3} Ch. 5). We consider the quantum supergroup $\rSL_q(4|1)$ in Manin's formalism \cite{ma}. In this section we use the same letters for the classical and quantum generators while it doesn't lead to confusion. We have the following definition:

 \begin{definition} \label{ManinCR} The  quantum matrix superalgebra $\rM_q(m|n)$ is defined as
$$
\rM_q(m|n)=_{\mathrm{def}}\C_q \langle z_{ij},\xi_{kl}\rangle/\cI_M
$$
where $\C_q\langle z_{ij},\xi_{kl}\rangle$ denotes the free
superalgebra over $\C_q=\C[q,q^{-1}]$
generated by the even variables
$$z_{ij},\qquad  \hbox{ for }\quad 1 \leq i,j \leq m \quad \hbox{ or } \qquad m+1 \leq i,j \leq m+n.$$
and by the odd variables
\begin{align*}&\xi_{kl}&&  \hbox{for  }\quad 1 \leq k \leq m, \quad m+1 \leq l \leq m+n \\&&&\hbox{or   }\,  m+1 \leq k \leq m+n, \quad 1 \leq l \leq m,\end{align*}
satisfying the relations $\xi_{kl}^2=0$. $\cI_M$ is an ideal generated by relations that we will describe shortly. We can visualize the generators as a matrix
\beq\begin{pmatrix}z_{m\times m}&\xi_{m\times n}\\
\xi_{n\times m}&z_{n\times n}\end{pmatrix}\,.\label{generatorsmatrix}\eeq

To simplify the notation it is convenient sometimes to have a common notation for even and odd variables.
$$
a_{ij}=\begin{cases} z_{ij} & 1 \leq i,j \leq m, \, \hbox{   or   } \quad
                         m+1 \leq i,j \leq m+n \\ \\
              \xi_{ij} &  1 \leq i \leq m,\quad  m+1 \leq j \leq m+n, \, \hbox{   or } \\&
                      m+1 \leq i \leq m+n, \quad 1 \leq j \leq m
\end{cases}
$$

 We  assign a parity to the indices $p(i)=0$ if $1 \leq i \leq m$ and  $p(i)=1$ otherwise. Then the parity of $a_{ij}$ is  $\pi(a_{ij})=p(i)+p(j)$ mod 2. Then the ideal $\cI_M$ is generated by the relations \cite{ma}:

\begin{align*}
&a_{ij}a_{il}=(-1)^{\pi(a_{ij})\pi(a_{il})}
q^{(-1)^{p(i)+1}}a_{il}a_{ij}, && \hbox{for  } j < l \\&&& \\
&a_{ij}a_{kj}=(-1)^{\pi(a_{ij})\pi(a_{kj})}
q^{(-1)^{p(j)+1}}a_{kj}a_{ij}, && \hbox{for  } i < k \\ &&&\\
&a_{ij}a_{kl}=(-1)^{\pi(a_{ij})\pi(a_{kl})}a_{kl}a_{ij}, &&  \hbox{for  }
i< k,\;j > l \\&&&\hbox{or } \quad i > k,\; j < l \\&&& \\
&a_{ij}a_{kl}-(-1)^{\pi(a_{ij})\pi(a_{kl})}a_{kl}a_{ij}=(-1)^{\pi(a_{ij})\pi(a_{kj})}(q^{-1}-q)
a_{kj}a_{il},&&\\
&&& \hbox{for  }\quad i<k,\;j<l
\end{align*}

$\hbs$
\end{definition}

There is also a comultiplication
$$\begin{CD}\rM_q(m|n)@>\Delta>>\rM_q(m|n)\otimes \rM_q(m|n)\end{CD}$$ given formally by matrix multiplication, that is,
$\Delta( a_{ij})=\sum_ka_{ik}\otimes a_{kj}.$

Adjoining the inverse of the quantum Berezinian, which is a central element of the algebra one has a  suitable antipode map $S$, unit and a counit that define the  standard quantum supergroup $\rGL_q(4|1)$. One can restrict to $\rSL_q(4|1)$ by setting the quantum Berezinian to 1. We represent conveniently the generators in matrix form as in (\ref{generatorsmatrix})
\beq\begin{pmatrix}g_{ij}&\gamma_{i5}\\\gamma_{5j}&g_{55}\end{pmatrix},\qquad  i,j=1,\dots ,4\,.\label{GLq41}\eeq
As in the classical case, the quantum super Poincar\'{e} group is given by
$$
\cO_q(P):=\cO(\rGL_q(4|1))/\cI_q\,,$$
where $\cI_q$ is the (two-sided) ideal in $\cO(\rGL_q(4|1))$
generated by
\beq g_{1j},g_{2j}, \quad \hbox{for}\quad j=3,4
\quad \hbox{and}\quad \gamma_{15},\gamma_{25}, \gamma_{53},\gamma_{54} \,.
\label{qsp}\eeq
We can write equivalently $\cO_q(P)$ as generated, as in (\ref{lpss}), by
the following elements:
\beq
\begin{pmatrix}x&0&0\\
 Tx&y& y\eta\\ \rho x&0&d\end{pmatrix}\,,\label{qps} \eeq
   with change of variables
  \begin{align*}
&x =\begin{pmatrix}
g_{11} & g_{12}\\ g_{21} & g_{22}  \end{pmatrix},
&&
T =
\begin{pmatrix} -q^{-1}D_{23}D_{12}^{-1} & D_{13}D_{12}^{-1}\\
-q^{-1}D_{24}D_{12}^{-1} & D_{14}D_{12}^{-1}  \end{pmatrix},
\nonumber \\\nonumber \\
&y = \begin{pmatrix}
g_{33} & g_{34}\\ g_{43} & g_{44} \end{pmatrix},&&
d =g_{55}, \\ \\
&
\rho=(  -q^{-1}D_{25}D_{12}^{-1},
D_{15}D_{12}^{-1} ),&& \eta =
\begin{pmatrix}
-q^{-1}{D^{34}_{34}}^{-1}D^{45}_{34} \\
{D^{34}_{34}}^{-1}D_{34}^{35} \\
\end{pmatrix}\,.
\end{align*}
We have  denoted by $D_{ij}$ the determinant of the $i, j$ rows of the first
two columns in (\ref{qsp}) (see \cite{fl3} Ch. 5 for more details).

In the following it will be necessary to distinguish between
(super) commutative and non (super)commutative generators, so we put a hat `$\,\hat{\phantom{o}}\,$' over the non commutative ones. In the spirit of (\ref{minkowskiss}) we introduce the following definition:
\begin{definition}\label{cqms}The {\it complexified quantum Minkowski superspace} is the free algebra in six generators
\begin{align*}
&\hv_{41} , \,\hv_{42} , \,\hv_{31} \, \hbox{ and } \,\hv_{32},\qquad &\hbox{(even)}\\
&\hu_{51}, \hu_{52} &\hbox{(odd)}
\end{align*}
satisfying the commutation relations

\begin{align}
&\hv_{42} \hv_{41}  =  q^{-1} \hv_{41} \hv_{42}, \nonumber\\
&\hv_{31} \hv_{41}  =  q^{-1} \hv_{41} \hv_{31}, \nonumber\\
& \hv_{32} \hv_{41}  =   \hv_{41} \hv_{32} + (q^{-1} -q ) \hv_{42} \hv_{31},\nonumber\\
& \hv_{31} \hv_{42}  =  \hv_{42} \hv_{31}, \nonumber\\
& \hv_{32} \hv_{42}  =  q^{-1} \hv_{42} \hv_{32}, \nonumber\\
& \hv_{32} \hv_{31}  =  q^{-1} \hv_{31} \hv_{32}\,, \label{Relations QM}
\end{align}
which would be the commutation relations defining the non super quantum Minkowski space,  together with

\begin{align}
&\hu_{51}\hu_{52}=-q^{-1}\hu_{52}\hu_{51}&& \nonumber\\
&\hv_{31}\hu_{51}=q^{-1}\hu_{51}\hv_{31}
&&\hv_{32}\hu_{52}=q^{-1}\hu_{52}\hv_{32}\nonumber\\
&\hv_{41}\hu_{51}=q^{-1}\hu_{51}\hv_{41},
&&\hv_{42}\hu_{52}=q^{-1}\hu_{52}\hv_{42}\nonumber\\
&\hv_{31}\hu_{52}=\hu_{52}\hv_{31}
&&\hv_{41}\hu_{52}=\hu_{52}\hv_{41}\nonumber\\
&\hv_{32}\hu_{51}- \hu_{51}\hv_{32}=(q-q^{-1})\hv_{31}\hu_{52}
&&\hv_{42}\hu_{51}- \hu_{51}\hv_{42}=(q-q^{-1})\hv_{41}\hu_{52}
\label{Relations QSM}
\end{align}

This algebra will be denoted as $\cO_q(\cM)$, and it is a subalgebra of $\rM_q(2|1)$ in Definition \ref{ManinCR}.
If we denote the ideal given by (\ref{Relations QM}) and
(\ref{Relations QSM}) as $\cI_{q\cM}$, then we have that
$$\cO_q(\cM)\equiv \C_q\langle \hv_{41}, \hv_{42}, \hv_{31}, \hv_{32}, \hu_{51},\hu_{52}\rangle/\cI_{q\cM}\,.$$

\hfill$\blacksquare$
\end{definition}

The commutation relations of the generators of the quantum Poincar\'{e} supergroup $\cO(P_q)$ {$\cO_q(P)$ }(\ref{qsp}), (\ref{qps}), are not trivial and are listed in \cite{fl3} (pages 305-306) in a very similar notation. There is a natural coaction of $\cO_q(P)$ on
$\cO_q(\cM)$ which is  Proposition  7.5 of \cite{cfl1}. We give it here:

\begin{proposition}
The quantum chiral super Minkowski space $\cO_q(\cM)$ admits a coaction of
the quantum chiral Poincar\'e supergroup  { $\cO_q(P)$}:
$$
\begin{CD}\cO_q(\cM)@>\hat\Delta>>  \cO_q(P) \otimes \cO_q(\cM)\,.\end{CD}$$
\begin{align*}
\hat\Delta \hat t_{ij}&=t_{ij}\otimes \id+\hat y_{ia} S(\hat x)_{bj}\otimes \hat t_{ab}+\hat y_{i}\hat\eta_aS(\hat x)_{bj}\otimes \hat\rho_{jb},&
\\ \\
\hat\Delta\hat\tau_j&=(\hat d\otimes \id)(\hat\tau_a\otimes \id+\id\otimes \hat\rho_a)(S(\hat x)_{aj}\otimes 1)\,.&
\end{align*}

\hfill$\blacksquare$
\end{proposition}

\section{The star product in the even case} \label{even-sec}

The star product allows us to recover the interpretation of the quantum algebra as the space of formal power series of  standard polynomials, where a non commutative product is defined. For the even case, the star product was computed in \cite{cfln}. We will make later the generalization to the super case, but we need  first to recall the construction in the non super case.

 \begin{definition}\label{cqmss}
The {\it complexified quantum Minkowski space} is the free algebra in four generators
$$
\hv_{41} , \hv_{42} , \hv_{31} \, \hbox{ and } \, \hv_{32}\,,
$$
satisfying the relations (\ref{Relations QM}).

This algebra will be denoted as $\cO_q(\cM_0)$. If we denote the ideal (\ref{Relations QM}) by $\cI_{\cM_{0q}}$, then we have that
$$\cO_q(\cM_0)\equiv \C_q\langle \hv_{41}, \hv_{42}, \hv_{31}, \hv_{32}\rangle/\cI_{\cM_{0q}}\,.$$

\hfill$\blacksquare$
\end{definition}


Let $\C_q=\C[q, q^{-1}]$. We have:

\begin{proposition}\label{moduleiso} There is an isomorphism
 $\cO(\cM_0)[q,q^{-1}]\approx \cO_q(\cM_0)$
 as modules over $\C_q$. In fact, the map
\be\begin{CD}\C_q[t_{41}, t_{42},t_{31}, t_{32}]@> Q_{\cM_0}>>\cO_q(\cM_0)\\t_{41}^a t_{42}^b t_{31}^c  t_{32}^d @>>> \hv_{41}^a \hv_{42}^b  \hv_{31}^c  \hv_{32}^d\end{CD}\label{QM}\ee is a module isomorphism (so it has an inverse).
\end{proposition}

{\sl Proof.} The ordering used here is the {\it normal ordering} given in  \cite{ma}. The Theorem 1.14 in this reference says that such monomials are a basis for the algebra generated by the entries of the  quantum matrix
$$\begin{pmatrix}
\hat t_{32}&\hat t_{31}\\
\hat t_{42}&\hat t_{41}
\end{pmatrix} $$
with commutation relations the Manin relations (\ref{ManinCR}) in the purely even, $n=2$ case.
 The proof of the proposition then follows.

\hfill$\blacksquare$

 A map like $Q_{\cM_0}$ is called an {\it ordering rule} or {\it quantization map}. In particular, Proposition \ref{moduleiso} is telling us that   {$\cO_q(\cM_0)$}is a free module over $\C_q$, with basis the set of standard monomials.

We can  pull back the product on {$\cO_q(\cM_0)$}  to  $\cO(\cM_0)[q,q^{-1}]$.

\begin{definition} The {\it star product} on $\cO(\cM_0)[q,q^{-1}]$ is defined  as
 \be f\star_{\mathrm{even}} g=Q_{\cM_0}^{-1}\bigl(Q_{\cM_0}(f)Q_{\cM_0}(g)\bigr),\qquad f,g\in \cO(\cM_0)[q,q^{-1}].\label{starprodM0}\ee\hfill$\blacksquare$
 \end{definition}

 By construction, the star product satisfies associativity. The algebra $(\cO(\cM_0)[q,q^{-1}], \,\star\,)$ is then isomorphic to $\cO_q(\M)$.

To give it explicitly \cite{cfln} we need first a couple of partial results.
We begin by computing some  auxiliary relations

\begin{lemma} The following commutation rules are satisfied in{ $\cO_q(\cM_0)$}:\label{lemma}
\begin{align*}
&\hv_{42}^m \hv_{41}^n = q^{-mn} \hv_{41}^n \hv_{42}^m, \\
& \hv_{31}^m \hv_{41}^n = q^{-mn} \hv_{41}^n \hv_{31}^m, \\
& \hv_{31}^m \hv_{42}^n = \hv_{42}^n \hv_{31}^m, \\
& \hv_{32}^m \hv_{42}^n = q^{-mn} \hv_{42}^n  \hv_{32}^m, \\
& \hv_{32}^m \hv_{31}^n = q^{-mn} \hv_{31}^n \hv_{32}^m,
\end{align*}
and
$$
\hv_{32}^m \hv_{41}^n = \hv_{41}^n \hv_{32}^m + \sum_{ k = 1}^{\mu} F_k(q, m,n) \hv_{41}^{n-k} \hv_{42}^k \hv_{31}^k \hv_{32}^{m-k},
$$
where {$\mu=\mathrm{min}(m, n)$. The coefficients}
$$
F_k(q,m,n) = \beta_k(q,m) \prod_{l = 0 }^{k-1} F(q , n - l )\hbox{ with }
F(q,n) = \left(\frac{1}{q^{2 n - 1}} -q\right)\label{efe}$$
and $ \beta_k(q,m) $ is defined by the recursive relation
$$
\beta_0(q,m)= \beta_m(q,m) = 1, \quad  \hbox{and}\quad \beta_k(q,m+1) = \beta_{k-1}(q,m) + \beta_k(q,m) q^{-2 k}.
$$
Moreover, $\beta_k(q,m)=0$ if $k<0$ or if $k>m$.
\end{lemma}

{\sl Proof}. The proof is just a (lengthy) computation. \hfill$\blacksquare$

\begin{theorem}\label{explicitstar} The star product given in Definition \ref{starprodM0} is given on two arbitrary monomials as
\begin{align}
&(t_{41}^a t_{42}^b t_{31}^c t_{32}^d )\star_{\mathrm{even}} (t_{41}^m t_{42}^n t_{31}^p t_{32}^r)  =  q^{-mc-mb-nd-dp} t_{41}^{a+m} t_{42}^{b+n} t_{31}^{c+p} t_{32}^{d+r} \; + \nonumber\\ & \sum_{k = 1}^{\mu=min(d,m)} q^{-(m-k)c -(m-k)b - n(d-k)-p(d-k)}  F_k(q,d,m)\cdot \; \\&t_{41}^{a+m-k} t_{42}^{b+k+n} t_{31}^{c+k+p} t_{32}^{d-k+r}\label{starpoly}
\end{align}

\hfill$\blacksquare$

\end{theorem}

We consider now a change in the parameter, $q=\exp h$. The classic limit is then obtained as $h\rightarrow 0$. One can expand (\ref{starpoly}) in powers of $h$. In \cite{cfln} it is  shown that, at each order in $h$, the star product  can be written as a bidifferential operator. Then the extension of the star product to $C^\infty$ functions is unique.

It is interesting to compute the antisymmetrization of the term of order 1 in $h$, which is the Poisson bracket

\begin{align}
\{f,g \}_\mathrm{even}=  &t_{41} t_{31} (\partial_{41}f \partial_{31}g-\partial_{41}g \partial_{31}f)  + t_{42} t_{41} (\partial_{41} f \partial_{42} g-
\partial_{41} g \partial_{42} f)+\nonumber\\& t_{32} t_{42} (\partial_{42}f  \partial_{32} g-\partial_{42}g  \partial_{32} f)+
    t_{32} t_{31} (\partial_{31} f\partial_{32}g-\partial_{31} g\partial_{32}f)  + \nonumber\\&2 t_{42} t_{31}( \partial_{41}f \partial_{32}g-\partial_{41}g  \partial_{32}f )\,.
\label{pbt}
\end{align}
We can express the Poisson bracket in terms of the usual variables in Minkowski space. The coordinate change is
$$\begin{pmatrix}t_{31}&t_{32}\\t_{41}&t_{42}\end{pmatrix}= x^\mu \sigma_\mu=\begin{pmatrix} x^0 + x^3 & x^1 - \ri x^2 \\ x^1 + \ri x^2 & x^0 - x^3 \end{pmatrix},$$ and the inverse change is
$$x^{0}=\frac{1}{2}(t_{31}+t_{42}),\quad
x^{1}=\frac{1}{2}(t_{32}+t_{41}),\quad
x^{2}=\frac{i}{2}(t_{32}-t_{41}),\quad
x^{3}=\frac{1}{2}(t_{31}-t_{42}).$$
In these variables the Poisson bracket becomes
\begin{align*}
\{f,g \}_\mathrm{even}= &\ri\Big(\left((x^{0})^{2}-(x^{3})^{2}\right)(\partial_{1}f\partial_{2}g -\partial_{1}g\partial_{2}f)  + x^{0}x^{1}(\partial_{0}f\partial_{2}g -\partial_{0}g\partial_{2}f)-\nonumber \\&
x^{0}x^{2}(\partial_{0}f\partial_{1}g
-\partial_{0}g\partial_{1}f)-x^{1}x^{3}(\partial_{2}f\partial_{3}g -\partial_{2}g\partial_{3}f)
+\nonumber\\&x^{2}x^{3}(\partial_{1}f\partial_{3}g -\partial_{1}g\partial_{3}f)\Big)\,.
\label{pb2}
\end{align*}
The Poisson bracket that we obtain is quadratic in the variables, and it is non trivial.

\section{The super star product}\label{superstarproduct-section}

We want to repeat the procedure of Section \ref{even-sec}. We start by giving an ordering which defines a quantization map.

\begin{proposition}\label{moduleiso2} There is an isomorphism
 $$\cO(\cM)[q,q^{-1}]=\C_q[\tau_{51},\tau_{52},t_{41}, t_{42}, t_{31},  t_{32} ]\approx \cO_q(\cM)$$
 as modules over $\C_q$. In fact, for $a,b,c,d=0,1,2\dots$ and $e,f=0,1$ the map
\beq\begin{CD}\cO(\cM)[q, q^{-1}] @> Q_{\cM}>>\cO_q(\cM)\\\tau_{51}^e\tau_{52}^ft_{41}^a t_{42}^b t_{31}^c  t_{32}^d @>>> \hu_{51}^e\hu_{52}^f\hv_{41}^a \hv_{42}^b  \hv_{31}^c  \hv_{32}^d\end{CD}\label{QSM}\eeq is a module isomorphism (so it has an inverse).
\end{proposition}

{\sl Proof.} The ordering used here is  the  {\it normal ordering} used in Theorem 1.14 of \cite{ma}.

For the even part we are in the situation of
Proposition (\ref{moduleiso}). Adding the odd variables  in that reference says that such monomials are a basis for the algebra generated by the entries of the (non square) quantum matrix
$$\begin{pmatrix}
\hv_{32}&\hv_{31}\\
\hv_{42}&\hv_{41}\\
\hu_{52}&\hu_{51}
\end{pmatrix} $$
with commutation relations the Manin relations (\ref{ManinCR}). To add the odd variables one has to check that they do not introduce other generators with commutation relations the Manin relations (\ref{ManinCR}).
These relations are the same (up to a sign) for even or odd generators.
Hence the normal ordering in Theorem 1.14 of \cite{ma}, will give us
the result using the same argument as in Prop. \ref{moduleiso}.

\hfill$\blacksquare$

We need first a couple of partial results
\begin{lemma}For $a, b,c, d=0,1,2,\dots $ we have the following commutation relations:
\begin{align*}
&\hat t_{41}^a\hat \tau_{52}=q^{-a}\hat \tau_{52}\hat t_{41}^a &
&\hat t_{42}^b\hat \tau_{52}=q^{-b}\hat \tau_{52}\hat t_{42}^b\\
&\hat t_{31}^c\hat \tau_{52}=\hat \tau_{52}\hat t_{31}^c &
&\hat t_{32}^d\hat \tau_{52}=q^{-d}\hat \tau_{52}\hat t_{32}^d\\
&\hat t_{41}^a\hat \tau_{51}=q^{-a}\hat \tau_{51}\hat t_{41}^a &\\
&\hat t_{42}^b\hat \tau_{51}=\hat \tau_{51}\hat t_{42}^b+(q-q^{-2b+1})\tau_{52}t_{41}t_{42}^{b-1}\\
&\hat t_{31}^c\hat \tau_{51}=q^{-c}\hat \tau_{51}\hat t_{31}^c &\\
&\hat t_{32}^d\hat \tau_{51}=\hat \tau_{51}\hat t_{32}^d+(q-q^{-2d+1})\tau_{52}t_{31}t_{32}^{d-1}\\
\end{align*}
\hfill $\blacksquare$

\end{lemma}

\begin{lemma}\label{lemma2}
For $a,b,c,d=0,1,2,\dots$
\begin{align*}
&t_{41}^a t_{42}^bt_{31}^ct_{32}^d\tau_{51}=q^{-(a+c)}\tau_{51}t_{41}^a t_{42}^bt_{31}^ct_{32}^d+ q^{-(a+c)}(q-q^{-2b+1})\tau_{52}t_{41}^{a+1} t_{42}^{b-1}t_{31}^{c}t_{32}^d+\\
&q^{-(a+b)}(q-q^{-2d+1})\tau_{52}t_{41}^a t_{42}^bt_{31}^{c+1}t_{32}^{d-1}\\&\\&
t_{41}^a t_{42}^bt_{31}^ct_{32}^d\tau_{52}=q^{-(a+b+d)}\tau_{52}t_{41}^a t_{42}^bt_{31}^ct_{32}^d
\end{align*}

\hfill $\blacksquare$

\end{lemma}
In order to simplify the notation we define
$$
T(a,b,c,d)=t_{41}^a t_{42}^bt_{31}^ct_{32}^d\,.
$$
As in the non super case we have
\begin{definition} The {\it super star product} on $\cO(\cM_0)[q,q^{-1}]$ is defined  as
 \be f\star g=Q_{\cM}^{-1}\bigl(Q_{\cM}(f)Q_{\cM}(g)\bigr),\qquad f,g\in \cO(\cM)[q,q^{-1}].\label{starprodM}\ee\hfill$\blacksquare$
 \end{definition}
 \begin{theorem}The super star product  of two monomials in the given basis
$$ S=\tau_{51}^e\tau_{52}^fT(a,b,c,d)\star\tau_{51}^u\tau_{52}^vT(m,n,p,r)\,,$$
with $e,f,u,v=0,1$ and $a,b,c,d,m,n,p,r=0,1,2,\dots$ is given, in terms of the even star product by (\ref{starprodM0}) (the exponents of the odd variables are always taken  mod(2), so they take values 0 or 1)
\begin{align}S=&\delta_{u0}\delta_{v0}\tau_{51}^e\tau_{52}^f\,T(a,b,c,d)\star_\mathrm{even} T(m,n,p,r)+\nonumber\\&
\delta_{u0}\delta_{v1}\left(q^{-(a+b+d)}\tau_{51}^e\tau_{52}^{f+1}\,T(a,b,c,d)\star_\mathrm{even} T(m,n,p,r)\right)\\&\delta_{u1}\delta_{v0}\left((-1)^fq^{f-a-c}\tau_{51}^{e+1}\tau_{52}^f\, T(a,b,c,d)\star_\mathrm{even} T(m,n,p,r)+\right.\\&q^{-a-c}(q-q^{-2b+1})\tau_{51}^e\tau_{52}^{f+1}\,T(a+1,b-1,c,d)\star_\mathrm{even} T(m,n,p,r)+\nonumber\\
&\left. q^{-a-b}(q-q^{-2d+1})\tau_{51}^e\tau_{52}^{f+1}\,T(a,b,c+1,d-1)\star_\mathrm{even} T(m,n,p,r)
\right)+
\nonumber\\&\delta_{u1}\delta_{v1}\left((-1)^fq^{-2a-b-c-d+f}\tau_{51}^{e+1}\tau_{52}^{f+1}\,T(a,b,c,d)\star_\mathrm{even} T(m,n,p,r)+\right.\nonumber\\& q^{-2a-b-c-d+1)}(q-q^{-2b+1})
\tau_{51}^{e}\tau_{52}^{f}\,T(a+1,b-1,c,d)\star_\mathrm{even} T(m,n,p,r)+\nonumber\\
&\left.q^{-2a-2b-d+1}(q-q^{-2d+1})
\tau_{51}^{e}\tau_{52}^{f}\,T(a,b,c+1,d-1)\star_\mathrm{even} T(m,n,p,r)
\right)  \,,\label{superstarproduct}
\end{align}
where the star products between the $T$'s are given in Theorem \ref{explicitstar}.
It is a quantum deformation of $\cO(\cM)$.

\end{theorem}

{\it Proof}. For the  proof we have to reorder according to \ref{lemma2}. It is easy to check that for $q=1$ one obtains the standard, (super) commutative product on $\cO(\cM)$.

\hfill$\blacksquare$

\bigskip The antisymmetrization of the first order in $h$ of the star product gives the Poisson bracket. To make the notation even lighter we will write
\begin{align*} &R_A=\tau_{51}^e\tau_{52}^ft_{41}^a t_{42}^bt_{31}^ct_{32}^d,\qquad  &&R_M=\tau_{51}^u\tau_{52}^vt_{41}^m t_{42}^nt_{31}^rt_{32}^p\,,\noindent \\
&R_A^+=\tau_{51}^e\tau_{52}^ft_{41}^{a+1} t_{42}^{b-1}t_{31}^ct_{32}^d,\qquad &&R_A^-=\tau_{51}^e\tau_{52}^ft_{41}^{a} t_{42}^{b}t_{31}^{c+1}t_{32}^{d-1}\,,\end{align*}

 when there is no possibility of confusion.
 Let us denote $S_1(R_A, R_M)$ the term of order $h$ in (\ref{superstarproduct}). We denote by $C_1$ the same term in (\ref{starprodM0}).

\begin{align}S_1(R_A,R_M)=&C_1(R_A,R_M)+\delta_{u0}\delta_{v1}\left(-(a+b+d)R_AR_M\right)\nonumber\\
&\delta_{u1}\delta_{v0}\left((f-a-c)R_AR_M+2bR_{A}^+R_M+2dR_{A}^-R_M\right)\nonumber\\
&\delta_{u1}\delta_{v1}\left((-2a-b-c+f)R_AR_M+2bR_{A}^+R_M+2dR_{A}^-R_M\right)\,.\label{semistar}
\end{align}

We now take into account that for $u,v=0,1$
\begin{align*}&\delta_{u0}=1-u,&&\delta_{v0}=1-v,\\
&\delta_{u1}=u,&&\delta_{v1}=v\,.
\end{align*}
Then the terms proportional to $uv$ in (\ref{semistar}) cancel out so we are left with
\begin{align*} S_1(R_A,R_M)=&C_1(R_A,R_M)-(a+b+d)R_A\cdot vR_M+(f-a-c)R_A\cdot uR_M+\\
&2bR_A^+\cdot uR_M+2dR_A^-\cdot uR_M
\label{semistarfinal}
\end{align*}

{$2bR_A^+\cdot uR_M+2dR_A^-\cdot uR_M$}

This expression can be written in terms of differential operators. We have, for example, that
$$ aR_A=t_{41}\partial_{t_{41}}R_A,\quad bR_A^+=t_{41}\partial_{t_{42}}R_A, \quad uR_M=\tau_{51}\partial_{\tau_{51}}R_M\,...$$
The result is
\begin{align*}S_1(R_A,R_M)=&C_1(R_A, R_M)+(t_{41}\partial_{ t_{41}}+t_{42}\partial {t_{42}}+t_{32}\partial _ {t_{32}})R_A\cdot\tau_{52}\partial_{\tau_{52}}R_M+\\&
(-\tau_{52}\partial_{\tau_{52}}-t_{41}\partial_{t_{41}}-t_{31}\partial_{t_{31}}+
2t_{41}\partial_{t_{42}}+2t_{31}\partial_{t_{32}})R_A\cdot\tau_{51}\partial_{\tau_{51}} R_M
\end{align*}
 which satisfies the Leibniz rule in both arguments. Its (anti)symmetrization is the Poisson bracket
$$\{R_A, R_M\}= S_1(R_A,R_M)-(-1)^{p_Ap_M}S_1(R_M,R_A)\,.$$
At this order the star product is differential. Presumably, it will be differential at all orders, as its non super counterpart \cite{cfln}.

\section*{Acknowledgments}

This work has been supported in part by grants  FIS2017-84440-C2-1-P  of the Ministerio de Econom\'{\i}a y Competitividad (Spain) and FEDER (EU) and  by the grant PROMETEO/2020/079 of the Generalitat Valenciana.

\appendix

\section{A basis for the Poincar\'{e} quantum supergroup}\label{diamond}

In this appendix we give a brief sketch on the fact that
the ordered monomials, according to a given ordering, of the
generators of the super Poincar\'{e} quantum group,
form a basis for its quantum superalgebra. This is a non trivial result
based on the classical work by G. Bergman                                        \cite{be}, but it is a
modification of the argument in \cite{cfln}.

We recall the following key result.

\begin{theorem} \label{diamondlemmathm} (Diamond Lemma).
Let $R$ be the ring defined by generators and relations as:
$$
R:=\C_q\langle x_i\rangle /(X_{I_k}-f_k,k=1 \dots s)
$$
If $\Pi=\{X_{I_k},f_k\}_{k=1, \dots, s}$ is compatible
with the ordering $<$ and all ambiguities are resolvable, then
the set of ordered monomials $\Pi$ is a basis for $R$. Hence $R$ is
a free module over $\C_q$.
\end{theorem}

Let us fix a total order $\cO$ on the variables $x$, $y$, $t$, $\tau$
as follows:
\begin{align*}
&\tau_{52}>\tau_{51}>t_{32} \, > \, t_{31} \, > \, t_{42} \, > \, t_{41}
\, > \\& x_{11} \, > \, x_{12} \, > \, x_{21} \, > \, x_{22} \, > \,
y_{33} \, > \, y_{34} \, > \, y_{43} \, > \, y_{44}.
\end{align*}


\begin{theorem} \label{diamlemmapoincare}
Let  $\cO_q(P)=\C_q\langle x_{ij},y_{kl}, t_{il}\rangle /\cI_{P}$
be the algebra corresponding to the quantum Poincar\'{e}
group. Then, the monomials
in the order $\cO$ as above
are a basis for $\cO_q(P)$.
\end{theorem}

The proof of this result is completely analogous to the non
super setting. It is an application of the Diamond Lemma, where
we resolve all ambiguities using the Manin relations,
whose form, in fact, does not depend on the   generators having parity.

\end{document}